# Recent Development in Analog Computation – A Brief Overview

Yang Xue

Department of Electrical Engineering and Computer Science

The University of Tennessee

Knoxville, TN 37996-2250, USA

yxue5@utk.edu

*Abstract* – The recent development in analog computation is reviewed in this paper. Analog computation was used in many applications where power and energy efficiency is of paramount importance. It is shown that by using innovative architecture and circuit design, analog computation systems can achieve much higher energy efficiency than their digital counterparts, as they are able to exploit the computational power inherent to the devices and physics. However, these systems do suffer from some disadvantages, such as lower accuracy and speed, and designers have come up with novel approaches to overcome them. The paper provides an overview of analog computation systems, from basic components such as memory and arithmetic elements, to architecture and system design.

I. INTRODUCTION

There are more and more application cases where the energy efficiency of computational elements is of paramount importance. One of such example is portable devices where the volume and weight of the battery is limited, and the power consumption needs to be minimized to extend the battery life. A more extreme case is the autonomous sensors where all the operation power is scavenged from environment [1]. In implantable biomedical devices, the power consumption has to be minimized so that the heat generated does not affect the nearby tissue, and battery life is maximized.

In order to accommodate this ever-increasing need for efficiency, researchers have designed the analog frontend of these systems with great performance and efficiency. For example, [2] [3] [4] discussed low-power high-performance frontend amplifiers and filters for implantable systems. Analog to digital converters (ADCs) [5] [6] [7] [8] [9] are usually an important and power-hungry part in these systems, and [10] [11] proposed comparator designs that can greatly improve the accuracy and efficiency of these ADCs. Computational units (processors) are another essential part, as they provide autonomous decision making, compression of sensory data, and other functions. These computational unit are usually implemented with digital processors, which can be too inefficient for ultra-low-power applications.

To explore the approaches that could lead to efficiency beyond digital processors, it is meaningful to compare the energy efficiency between the human brain and the digital processor. The human brain has an energy efficiency of about $10^{15}$ operation per joule [12]. Today's super computer's energy efficiency is about $8.3 \times 10^9$ operation per joule [13], more than 6 orders of magnitude lower than the human brain. The huge energy efficiency gap suggests that the ways they do computation are fundamentally different. One notable difference is how the information is presented. Digital computer uses binary system, that is, systems using only 0 and 1 to achieve better immunity to noise. Whereas the neurons are using a continuous presentation of information, more like analog circuits. For example the firing rates and the membrane potential in a neuron are continuous variables. These analog signaling implies that a single wire can carry multi-bit information, as a result, the wiring and signaling overhead can be largely reduced. In addition, neurological system cleverly exploits the physical phenomena as the bases for computation. Aggregation and integration are carried out in charge domain, using the conservation of charge. Summing is implemented by observing Kirchhoff's current law (KCL). Amplification and non-linear functions are realized by modulation of energy barriers to ions in thermal equilibrium and therefore whose energies are Boltzmann distributed [14]. All these principle of operation is actually found and available in integrated circuits [15]. However, by using only on and off states of active devices, digital computers lose a factor of $10^4$ in efficiency [14]. Analog circuits have principles that are much more similar to neurological systems and have the potential to get back the lost efficiency. However, the analog computation is for the same reason more sensitive to noise and offset when compared to digital computers. It is found in [16] that the cost for precision for analog system increases exponentially, while that for digital increases polynomially. Therefore, analog computation is more preferable at low accuracy. Interestingly, the brain is also known to be of low accuracy, with a typical signal-noise ratio (SNR) in the range of 1-10 [17], [18], [19]. This in turn suggests that very high energy efficiency can be achieved if the system is built to tolerate the relative inaccuracy of analog computation.

In this paper, we will discuss different aspects of analog computation system, starting from basic components such as memory and arithmetic elements, to the entire architecture design of an analog deep learning engine.

## II. ANALOG MEMORY

Memory is one of the indispensable components in a computation system. Most of modern digital memories are volatile, either requiring constant refreshing, or a minimum VDD for state retention [20]. Non-volatile digital memories including flash memory [21], FRAM (Ferroelectric RAM) [22], [23], ReRAM (Resistive RAM) [24], and MRAM (Magnetoresistive

RAM) [25] require special process and add to fabrication cost. In analog computation systems, digital memory has inherent difficulty in integration because they require analog-digital conversion to interface with other circuits in analog domain. Recently, researchers have proposed different designs of Floating gate (FG) memories to realize analog non-volatile storage of information.

A floating gate device has gate that is electrically isolated from outside (floating), and the charge trapped on this isolated gate is used to store analog or digital values. To modify the charge on the floating gate, two mechanisms can be used. When a high voltage is applied across the gate oxide, the oxide barrier becomes more triangular and its effective thickness is reduced, making it easier for the quantum tunneling of electrons from the gate to the channel to happen. This is called Fowler–Nordheim (FN) tunneling and is a major cause of gate leakage current in modern deep-sub-micron process and can be measured and characterized with specially designed circuits [26]. As the reverse process of FN tunneling, hot electron injection can be exploited to add electron to the FG. When the source-to-drain voltage of a PFET is sufficiently large, the energy of holes near the drain side is large enough to generate hot electron-hole pairs. The liberated hot electrons are able to overcome the oxide barrier and be attracted to the floating gate, which is at positive potential.

In contrast to people's perception that FG devices can only be used to build digital memories such as EEPROM of Flash memories, they are actually more of an analog device, as the amount of charge on the FG is change in a continuous way. In the literature, FG devices are extensively used to implement analog systems [27], [28], [29], [30], [31], [32].

One of the important design difficulties of FG analog memories is to achieve tunneling control. Because FN tunneling requires high voltage, it is relative more difficult to control as the high voltage needs to be applied on or removed from the gate selectively. Therefore, many works [30], [31] choose not to have individual control of tunneling, instead, they use tunneling as global erase, and injection to program individual memory to its target, making it difficult to be used in adaptive systems. Selective tunneling is proposed in [33], by simultaneous control of the tunneling and control gate voltages. This requires a large number of pins and therefore not suitable for large-scale systems. In [27], a high-voltage switch is proposed to control the tunneling voltage individually, which is not compatible with standard process and consumes power. A charge pump is used to generate a local HV for the selected memory [32].

The update rules describe how the memory output changes with the update command. Single-transistor memory [33] tends to have exponential update rules, which are not desirable for the stability of algorithm [32]. By keeping the FG voltage constant during update, a more linear update rule can be realized [27] [32].

Paper [34] presents floating-gate analog memory with current mode output. It uses a novel approach to realize random access of individual memory in an array, as well as updates at both direction (tunneling and injection), without the use of charge pump or high voltage switch. The interconnection and pin count is minimized, facilitate scaling. And the memory has a pseudo-linear sigmoid update rule, which is more preferable than exponential rules. The memory design is compatible with standard digital processes and was implemented in a 0.13μm digital CMOS process, achieving small area and low power consumption. The performance of the analog memory in [34] is summarized in Table I.

TABLE I. PERFORMANCES SUMMARY OF THE FLOATING GATE MEMORY [34]

| Parameter | Value |
| --- | --- |
| Technology | 1P8M 0.13-μm CMOS |
| Area | 35×14 μm$^2$ |
| Power supply | 3 V |
| Power consumption | 45 nW |
| Output range | 0 - 10 nA |
| Programming resolution | 7 bits |
| Dynamic range | 53.8 dB |
| Programming isolation | 86.5 dB |

III. ANALOG NON-LINEAR OPERATOR

One of the advantages of analog computation is that some non-linear functions can be realized directly, while in digital processor, they require lots of computational resources. One family of important non-linear functions is radial-basis functions. These bell-shaped functions are extensive utilized, because they are a good metric of similarity in artificial neural network and machine learning systems [30], [35], [36], [37]. A bump circuit [38] provides arguably the simplest way to generate this kind of function, although this implementation does not have tunable width of the transfer function.  In [30], a floating gate input stage is added to scale the input to get a variable width.  Researchers use a digital-to-analog converter (DAC) to for similar purpose [36], both of these two approaches increases area and power consumption. Paper [35], [39] propose to switch the number of transistors connected in the circuit to vary the bump width, which leads to limited number of achievable transfer functions. A direct synthesis method is presented in [40], where the squared difference is exponentiated to yield a Gaussian function; however, the circuit design is relative complex and occupies large area.

By combining the current correlator discussed in [38] and a tunable transconductor, paper [41] achieve both variable width and height of the bump function. In [41], the design of linear trans-conductors in sub-threshold CMOS is discussed. Such design is challenging because the linear range of differential pair becomes smaller as the device enters weak inversion [42]. Linearization techniques have been proposed for strong inversion [42], [43], [44] [45], but none of these is adequate with biasing current down to nA. In [41], the output resistance of saturated transistor is exploited to implement a tunable resistance with very large value, which is used as the degeneration resistor for the trans-conductor. A novel pseudo-differential structure is also

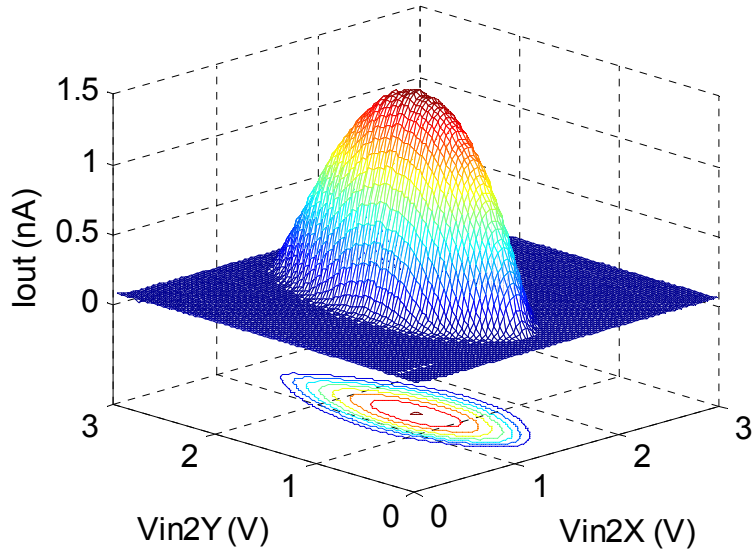

Figure 1: The measured 2-D bump output with different width on x and y dimensions [41].

Table II. Performance Summary and Comparison of the Bump Circuits

|  | [41] | [30] | [39] | [40]* |
|---|---|---|---|---|
| Technology | 0.13 μm | 0.5 μm | 0.13 μm | 0.18 μm |
| Supply voltage | 3 V | 3.3 V | 1.2 V | 0.7 V |
| Power | 18.9 nW | 90 μW | 10.5 μW | 485 nW |
| Area | 988 μm$^2$ | 3444 μm$^2$ | 1050 μm$^2$ | - |
| Response time | 45 μs | 10 μs | - | 9.6 μs |

*: simulation results

proposed to allow operation with supply voltage as low as 0.5V. In Figure 1, the measured 2-D bump output is plotted by cascading two identical bump circuits, with each dimension's parameters individually tunable. Table II summarizes and compares the performances of recently reported tunable bump circuits.

IV. ANALOG ONLINE CLUSTERING ENGINE

K-means clustering algorithm finds the underlying regularity in multi-dimensional data by capturing the spatial density patterns, and is used in many applications such as feature extraction and pattern recognition [46]. The algorithm is relative resource-intensive in digital domain, as it requires the parallel operation of multi-dimensional matrices. Analog computation has potential to improve upon digital processors in energy efficiency when implementing this algorithm [16]. Previous works have proposed analog vector quantizers (VQ), a simplified clustering engine, which search a set of pre-programmed templates and find the one that closest to the input [47], [48]. The clustering engine is one step further than VQ as it is able to adapt the templates according to the input statistics and therefore is more flexible in the face of changing environment. Analog clustering engines are relatively less reported due to their difficulty of design. The two designs in [49] and [36] have the template stored in volatile memories, precluding their use in applications with intermittent power. The use of digital memory in [36] requires analog to digital conversions for interfacing and adds to headroom. A novel analog online clustering engine is presented in [50]. Apart from the real-time online clustering capability, it is also capable of providing a measure of similarity between the input and each template based on simplified Mahalanobis distances, which greatly facilitates its integration to a more complex system. The learned templates are stored in a non-volatile fashion, enabling reuse after black-out. The prototype design with 8 by 4 dimensions was fabricated in a 0.13 μm digital CMOS process.

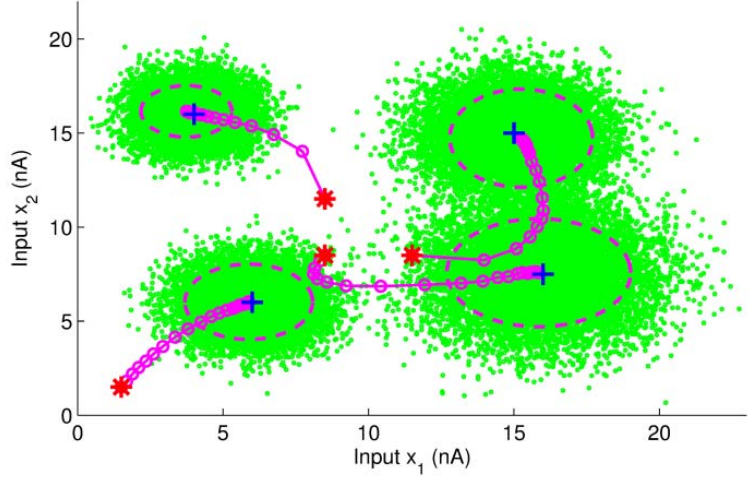

Figure 2: Clustering test result.

TABLE III. PERFORMANCE SUMMARY OF THE CLUSTERING CIRCUIT

| Parameter | Value |
| --- | --- |
| Technology | 1P8M 0.13 μm CMOS |
| Total Area | $0.9 \times 0.2$ mm$^2$ ($8 \times 4$ array) |
| MDC Cell Area | $90 \times 30$ μm$^2$ |
| Power consumption | 15 μW @ 3V |
| Classification Speed | 16 kHz |
| Clustering Speed | 4 kHz |

Figure 2 shows the clustering test results of the prototype [50]. The green dots are input vectors with 4 underlying clusters, each with different mean and sigma. The magenta lines show the evolution of template during clustering. Clearly the prototype was able to find the centroids and variances of the input data clusters accurately, demonstrating a robust learning performance. The measured performance is summarized in Table III.

## V. ANALOG MACHINE LEARNING ENGINE

The era of big data witnesses the extensive research and application of machine-learning [51]. Among the various subjects of machine learning, deep machine learning (DML) architectures mimic the information representation in the human brain, with the objective to achieve robust automated feature extraction, and provides a solution to the "curse of dimensionality" [52].

DML systems require massive parallel computation, which researchers have been realized with GPU-based platforms [53]. However these implementations are usually power hungry, making it hard to incorporate them to portable or wireless devices, and restrict their potential of scaling-up. Analog computation provides the potential to overcome this limit by achieving much higher energy efficiency compared to their digital counterpart.

Analog computation has been employed in machine learning systems and their building blocks [47] [36] [35] [30] [49] [41]. However, the ability to learn is what lacks in many of these designs. As a result, parameters have to be pre-programmed in order for the systems to function properly [47], [35], [30]. This places serious limitation on the application of these systems. A machine learning system with supervised learning capability is discussed in [36], but it still relies on the label provided during the training.

In these systems, memories are usually implemented in digital domain, which requires analog-digital conversions in order to interface with other analog circuitries. These additional conversions consume unnecessary area and power headroom, and require the memory to be centralized instead of distributed [54], [55], [56], [36]. In [49], analog values are stored on capacitors, this storage technique is inherently analog, but due to the leakage, it requires frequent refreshing, and the values drift over long term. These volatile storages are disadvantage because they will lose their states in the event of blackout, which can be common in battery-less applications where the power is harvested from the environment. [1].

In [37] [57], an analog deep learning engine is presented. It learns by unsupervised training, which is more preferable compared to supervised ones in that it does not require external assistant and therefore achieves fully autonomous operation. The learned parameters are stored in analog floating gate memories, which is non-volatile and compatible to standard digital processes. For easy scaling, the architecture adapts an array-based structure, which provides future possibility of large-scale implementation. The energy efficiency is the paramount design goal of this work and the authors uses several innovative design strategies to maximize the efficiency including parallel computation, deep sub-threshold biasing, algorithmic feedback to mitigate analog non-ideality [58], current mode arithmetic, distributed memory and flexible power domain partition. In Figure 3, image feature extraction is demonstrated, which reduces the input dimension from 32 to 4. The measured peak energy efficiency is over 1 Terra-OPS per Watt, and it is shown that the accuracy achieved is comparable to emulations in floating-point software environment, also shown in Figure 3. The performances of recent state-of-art parallel processors which employ bio-inspired analog computation are listed and compared in Table IV.

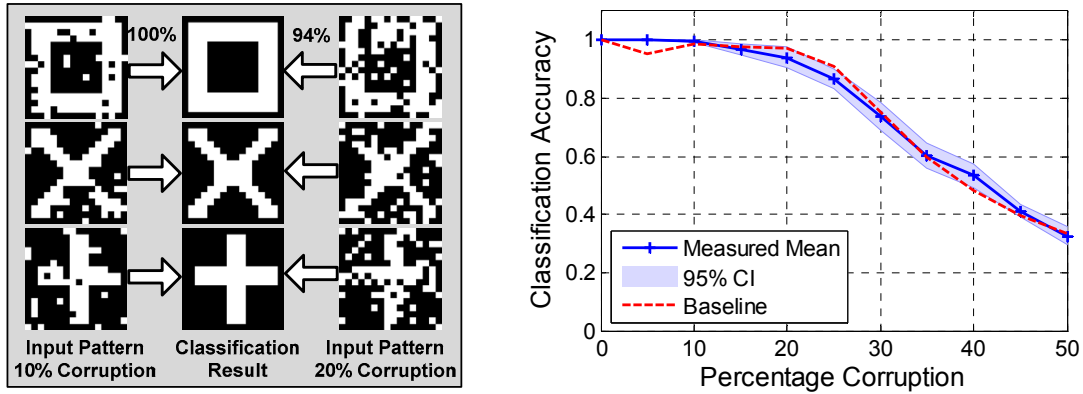

**Figure 3: Measured classification accuracy using the feature extracted by the chip [57].**

TABLE IV. COMPARISON OF ANALOG PROCESSORS

|  | JSSC'15 [57] | JSSC'13 [54] | ISSCC'13 [55] | JSSC'10 [56] |
|---|---|---|---|---|
| Process | 0.13um | 0.13um | 0.13um | 0.13um |
| Purpose | DML Feature Extraction | Neural-Fuzzy Processor | Object Recognition | Object Recognition |
| Non-volatile Memory | Floating Gate | NA | NA | NA |
| Power (W) | 11.4u | 57m | 260m | 496m |
| Peak Energy Efficiency | 1.04TOPS/W | 655GOPS/W | 646GOPS/W | 290GOPS/W |

## VI. Conclusions

This paper provides a brief survey of recent development in analog computation. Although analog computation has been around for a long time, researchers have recently focused on this topic as it shows the potential of energy efficiency improvement over digital processors. However, as analog systems require custom design and are sensitive to circuit non-idealities, innovative design and careful considerations on circuit, algorithm and architecture levels are important to achieve good performance.